\documentclass[aps,showpacs,twocolumn,prb]{revtex4-1}

\usepackage{graphicx}
\usepackage{amsmath}

\begin{document}

\title{Interlayer coupling in spin valves studied by broadband ferromagnetic resonance}
\author{D. E. Gonzalez-Chavez}
\author{R. Dutra}
\author{W. O. Rosa}
\author{T. L. Marcondes}
\author{A. Mello}
\author{R. L. Sommer}
\affiliation{Centro Brasileiro de Pesquisas F\'{\i}sicas,22290-180 Rio de Janeiro, RJ, Brazil}

\date{\today}

\begin{abstract}
Magnetization dynamic response of coupled and uncoupled spin valves with structure $\rm NiFe(20 nm)/Cu(t_{Cu})/NiFe(20 nm)/IrMn(10nm)$ is probed using broadband ferromagnetic resonance absorption measurements.
The coupling intensity between the free and pinned layers is tailored by varying the Cu thickness $\rm t_{Cu}$.
Broadband spectra exhibit two resonant modes for each value of applied field. It is observed that the coupling among NiFe layers modifies the amplitude of the absorption peaks and also the shape of the dispersion relations for each mode, which becomes particularly distorted at the anti-parallel magnetization state.
The observed phenomena is well described by applying a semianalytical model that properly takes into account the coupling interactions and allows an efficient numerical calculation of the absorption peak amplitudes, and the dispersion relation shapes.

\end{abstract}


\maketitle


\section{Introduction}

Interlayer coupling is an important ingredient for several devices as spin valves \cite{PhysRevB.43.1297, JPhysSocJpn.59.3061} and magnetic tunnel junctions \cite{PhysRevLett.74.3273, JMagnMagnMater.139.L231, Nat.Mater.3.862, Nat.Mater.3.868} (MTJ), multilayered materials and any systems based on two or more ferromagnetic layers separated by a nonmagnetic spacer.

On spin valves and MTJ, a strong interlayer coupling is a key issue for devices using synthetic free or pinned layers \cite{IEEETransMagn.32.4624, JApplPhys.83.3720, JApplPhys.87.5744}, while a weak coupling is usually observed between the free and pinned layers \cite{JApplPhys.76.1092, PhysRevLett.89.107206, ApplPhysLett.89.112503}.
In both cases the dynamical behavior is influenced by the strength of interlayer coupling, both, in saturated and not saturated magnetic states. 

This coupling was extensively studied in the past by several experimental techniques as magnetization measurements and magneto-resistance \cite{PhysRevLett.64.2304, PhysRevLett.67.3598}, ferromagnetic resonance \cite{PhysRevLett.63.1645, PhysRevB.44.9348, PhysRevB.47.5077, PhysRevB.71.224406}, Brillouin light scattering (BLS) \cite{PhysRevB.62.16109, PhysRevB.50.6143, JApplPhys.84.958}, and others \cite{PhysRevB.76.104414, PhysRevB.84.134406}. 

An interesting and recent approach to study the effect of interlayer coupling on the high frequency response is the use of broadband ferromagnetic resonance. This technique is based on the use of a vector network analyser (VNA), and is usually known as VNA-FMR \cite{IEEE.T.MTT.34.80}. Using this technique  we are able to measure the dynamic properties (permeability or absorption) in a frequency range from a few MHz to dozens of GHz. 
Moreover, all measurements can be performed in the range $- \rm H_{max} \leq 0 \leq + \rm H_{max}$, where $\rm H_{max}$ can be adjusted from a few Oe to several kOe. Therefore, besides measuring the saturated states as in traditional FMR, a broadband measurement can be performed on non saturated states, and eventually at zero field.

In this work, we study the static and dynamic properties of spin-valve systems using VNA-FMR and magnetometry measurements. Our samples are consisted of Py/Cu/Py/IrMn layers described as follows. The bottom Py = permalloy ($\rm Ni_{81}Fe_{19}$) layer acts as a free magnetic layer (F) while the top Py layer is coupled to an antiferromagnet ($\rm Ir_{20}Mn_{80}$) and behaves as a pinned layer (P). We are able to address the behavior of each layer and the effect of the interaction mediated by the Cu spacer. By  varying Cu layer thickness we are able to control the interaction between the Py layers, which produces new features on broadband spectra at non-saturated magnetic states. In particular, we observe complex dispersion relations, including frequency jumps and absorbed power intensities depending on the oscillation modes.  

A semianalytical model based on the magnetic free energy for the macro spins, together with the Landau Lifshitz Gilbert equation (LLG) is proposed and applied to these systems.
This model allows an efficient numerical calculation of the broadband absorption amplitudes and dispersion relations describing remarkably well the experimental results.
Moreover, the model provide further insights on the magnetization dynamics of spin-valve like systems in both, saturated and non-saturated magnetic states.

\section{Experiment}

We produced spin-valves with structure Py(20\,nm)/Cu($t_{ \rm Cu}$)/Py(20\,nm)/IrMn(15\,nm), where $t_{ \rm Cu}$ = 0.75\,nm, 1.0\,nm and 2.5\,nm, were produced using a Magnetron Sputtering system onto a Si(100) substrate with both buffer and capping layers of Ta(5\,nm). Chamber pressure condition for such depositions was 5\,mTorr/50\,sccm Ar pressure/flow, after a $5\times 10^{-8}$ Torr base pressure in the whole chamber. 
A RF power source was used for Py depositions, while DC sources were used for Ta, Cu and IrMn depositions. All deposition rates were calibrated using low angle x-ray reflectometry. During the growth process, an in-plane magnetic field of about 200\,Oe was applied in order to induce an unidirectional anisotropy at the FM/AFM interface, leading to the pinning of the top FM layer through exchange bias effect.

\begin{figure}[h!]
\includegraphics{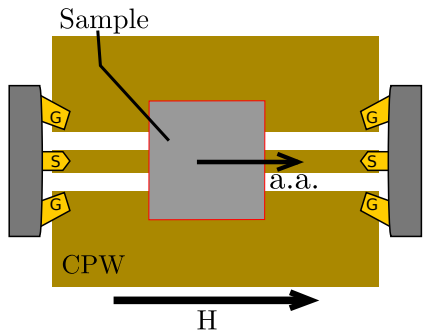}
 \caption {\label{fig:ExpSetup}  Schematic diagram of the coplanar waveguide (CPW) structure and a sample placed on top of it. The central conductor of the CPW is about 260\,$\mu$m wide. High-frequency micro-probes and coaxial cables (not shown) were used to connect the structure through VNA. The sample's anisotropy axis (a.a.) is aligned to the direction of the applied external field {\rm H}.}
\end{figure}

We performed static magnetic measurements ($M$ vs. $\rm H$) using a VSM under DC fields of $\pm$ 300\,Oe.
For the dynamic measurements, we used a broadband ferromagnetic resonance setup composed by a Rhode$\&$Shwarz ZVA24 Vector Network Analyser, combined with a coplanar waveguide for frequencies in the range of 0.1 - 7.0 GHz and DC magnetic fields in range of $\pm$ 300\,Oe. 
For these measurements, each sample was placed on top of a two port coplanar waveguide, where the external field $\rm H$ was applied along the propagation direction, as shown in Fig. \ref{fig:ExpSetup}. The transmission $S_{21}$ and reflection $S_{11}$ coefficients were measured in such specified field and frequency range.
The absorbed power ratio in the waveguide was calculated using \cite{IEEE.Trans.Magn.37.561}
\begin{equation}
P_\mathrm{Loss}/P_\mathrm{In} = 1 - |S_{11}|^2 - |S_{21}|^2.
\end{equation}
The ferromagnetic resonant spectra (magnetic absorption) were obtained by measuring this ratio with respect to a reference measurement of the dielectric losses, acquired with the sample saturated along the direction of the rf field. 

\section{Experimental Results}

\begin{figure}[h!]
\includegraphics{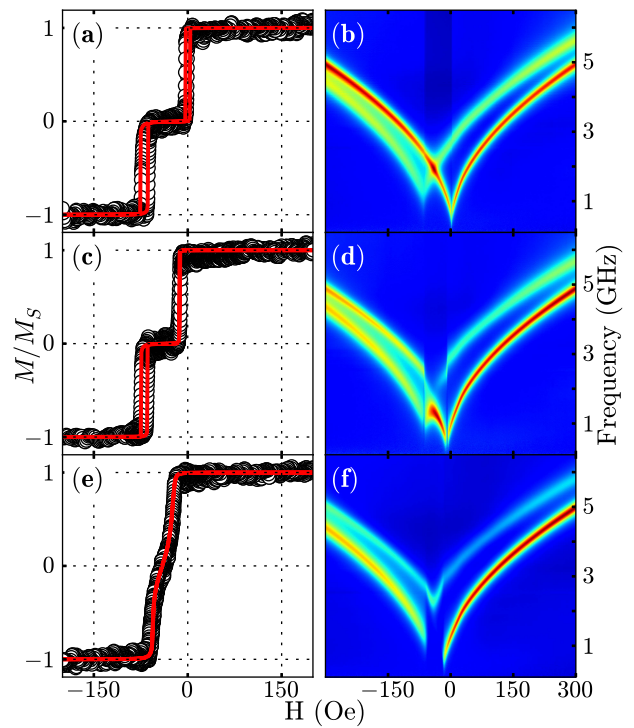}
 \caption {\label{fig:ExpMeas} Measured and calculated magnetic hysteresis loop (left) and the broadband FMR spectra (right) for $t_{\mathrm{Cu}}$ = 2.50\,nm (top) $t_{\mathrm{Cu}}$ = 1.00\,nm (middle) and $t_{\mathrm{Cu}}$ = 0.75\,nm (bottom). The symbols correspond to the experimental data and the solid line to the calculated curve.
 }
\end{figure}

Our samples were engineered in order to have different coupling intensities between the FM layers. 
From the static hysteresis loops, measured with the external field applied along the easy axes (as shown in Fig. \ref{fig:ExpMeas}), we can clearly note how the thickness of Cu spacer intermediates the intensity of the coupling among the FM layers.
The sample with $t_{\mathrm{Cu}}$ = 2.5 nm (Fig. \ref{fig:ExpMeas}(a)) has showed the typical spin-valve behavior, with well-known parallel and anti-parallel magnetization states, displaying a shifted response for P and a centered response for F layers. Such features indicate non appreciable coupling between the FM layers. On the other hand, for $t_{\mathrm{Cu}}$ = 1.00\,nm (Fig. \ref{fig:ExpMeas}(b)), the coupling now manifests itself as a small shift in the response of the F layer. A larger coupling is obtained for $t_{\mathrm{Cu}}$ = 0.75\,nm, where the shift of the F layer is larger and the anti-parallel state is no longer observed. Instead of that, a gradual rotation of the magnetization is actually the main switching process.

\begin{figure}[h!]
\includegraphics{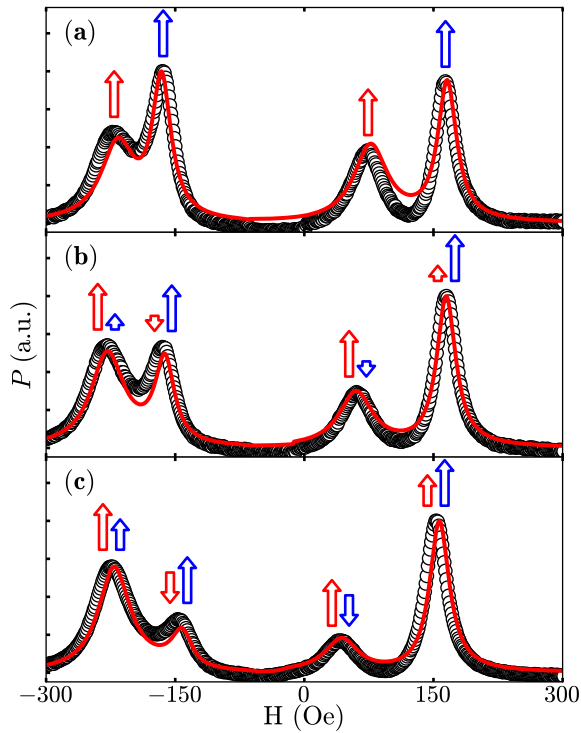}
 \caption {\label{fig:AbsPeaks} Absorbed power profiles at 3.7\,GHz for the samples with different Cu spacer thicknesses (a) $t_{\mathrm{Cu}}$ = 2.5\,nm, (b) $t_{\mathrm{Cu}}$ = 1.0\,nm and (c) $t_{\mathrm{Cu}}$ = 0.75\,nm. The symbols correspond to the experimental data and the solid line to the calculated curve.  Arrows represent the oscillating vectors, see Fig \ref{fig:Vectors} for further details.}
\end{figure}

The right side of Fig. \ref{fig:ExpMeas} shows the measured absorbed power spectra for our samples. The color scale denotes the amplitude from blue (minimum) to red (maximum). The maximum amplitude on branches correspond to the resonant modes. 
In these measurements, we are able to observe two clear resonant responses (Fig. \ref{fig:ExpMeas}(d)) for the sample without coupling ($t_{\mathrm{Cu}}$ = 2.50 nm), one centered and the other field shifted, corresponding to F and P layers, respectively. As already observed \cite{,JApplPhys.106.063918} in simple or exchange biased magnetic systems, the change in the slopes of the branches occurs at the switching fields of the respective layers.
On the other hand, for the samples with coupled FM layers ($t_{\mathrm{Cu}}$ = 1.00\,nm and $t_{\mathrm{Cu}}$ = 0.75\,nm), a pair of resonant branches is still observed in parallel magnetization states, while a completely different and a new behavior is observed in non-saturated states, including frequency jumps in the resonant branches for both layers at their switching fields. These features will be addressed in the following section \ref{sec:Discussion}, after we present our model and numerical calculations for these systems.
In all cases (saturated and non-saturated samples) we observe different absorption intensities on the resonant branches. In order to get a further insight on the absorption of the saturated states, we plot the absorption profile for these samples at 3.7 GHz in Fig.  \ref{fig:AbsPeaks}.
In this figure, four absortion peaks are observed for all samples. In the uncoupled case (Fig. \ref{fig:AbsPeaks}(a)), the small peaks correspond to the oscillation of the P layer, while the larger peaks are associated to the resonance of the F layer. Therefore, the difference in the height of peaks is clearly ascribed to the larger damping parameter $\alpha$ for the exchange biased P layer \cite{PhysRevB.58.8605}. 
For the samples with interaction between the FM layers (Fig. \ref{fig:AbsPeaks}(b-c)), the inner peak amplitudes decrease with respect to the outer peaks. Such decreasing seems to depend on the coupling intensity and it will be explained onward by our model and numerical calculations. 

\section{Semianalytical Model and Numerical Calculations}
\label{sec:model}
In order to understand the features observed in our $M$ vs. $\rm H$ curves and broadband measurements, we have adopted a macro spin model which takes into account the usual free energy density terms for each ferromagnetic layer plus a term describing the effective interaction between the free (F) and pinned (P) layers, as follows:
\begin{equation}
E = E_\mathrm{Pinned} +  E_\mathrm{Free} + E_\mathrm{Interaction}
\end{equation} 

\begin{figure}[h!]
\includegraphics{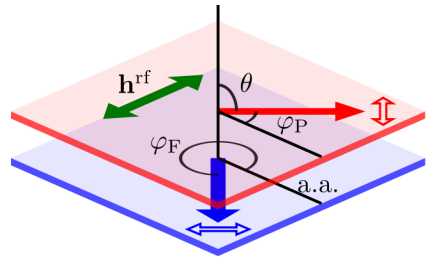}
 \caption {\label{fig:System}Schematic diagram of the theoretical system considered for the numerical calculations. The magnetization vectors (filled arrows) lay on the plane of the samples, their orientations are defined by the $\theta$ and $\varphi$ angles measured from the sample's normal and the anisotropy axis (a.a.), respectively. The oscillating vectors (empty arrows) are parallel to the $\hat{\varphi}$ directions (not show). The radio frequency field $\mathbf{h}^{\rm rf}$ is also parallel to the sample and perpendicular to a.a.}
\end{figure}

The free energy density for each layer $E_\mathrm{Pinned}$ and $E_\mathrm{Free}$ incorporate the Zeeman, in-plane uniaxial anisotropy, shape anisotropy and out-of-plane anisotropy terms.
$E_\mathrm{Pinned}$ includes also the exchange bias interaction term that keeps the corresponding layer pinned and the related rotatable anisotropy \cite{PhysRevB.58.8605}.
In our system, as shown in Fig. \ref{fig:System}, both layers have the same thickness $t$ and saturation magnetization $M_S$ and we express the energy density in terms of the polar $\theta$ and azimuthal $\varphi$ angles of the magnetizations and the anisotropy axes.
Since the shape anisotropy energy is the dominant in our system, the magnetization vector always lays on the plane of thin films, therefore $\theta = \pi/2$. Considering also that the anisotropies have an in-plane easy axis direction that is parallel to $\varphi = 0$ and having the external field ${\rm H}$ applied at $\varphi_{\rm H}$, then the normalized free energy density $(\eta = E/M_S)$ for F and P layers, keeping only the  $\varphi$ dependent terms, can be written as follows:
\begin{align}
\begin{split}
\eta_{\rm Pinned} = & - {\rm H} \cos(\varphi_{\rm H} - \varphi_{\rm P}) 
                      - {\rm H_{EB}} \cos(\varphi_{\rm P}) \\ 
                    & - \frac{1}{2} {\rm H_k^{P}} \cos^2(\varphi_{\rm P}) \\ 
\eta_{\rm Free} = & - {\rm H} \cos(\varphi_{\rm H} - \varphi_{\rm F})
                    - \frac{1}{2} {\rm H_k^{F}} \cos^2(\varphi_{\rm F})                   
\end{split}
\end{align}
where ${\rm H_k^{P}}$, ${\rm H_k^{F}}$, $\varphi_{\rm P}$ and $\varphi_{\rm F}$ are the uniaxial anisotropy fields and the in-plane magnetization angles of the P and F layers, respectively; $\rm H_{EB}$ is the exchange bias field acting on the pinned layer.
The adopted interaction energy density reads:
\begin{equation}
 \eta_\mathrm{Int} =  - {\rm H_1^J} \cos(\varphi_{\rm P} - \varphi_{\rm F}) 
 					  + {\rm H_2^J} \cos^2(\varphi_{\rm P} - \varphi_{\rm F}) 
\end{equation}
with ${\rm H_1^J} = \frac{\rm J_1}{t\,M_S}$ and ${\rm H_2^J} = \frac{\rm J_2}{t\,M_S}$, where ${\rm J_1}$ and ${\rm J_2}$ are the bilinear and biquadratic interaction constants between the two layers, respectively.

By minimizing $\eta = \eta_{\rm Pinned} + \eta_{\rm Free} + \eta_\mathrm{Int}$ for a given ${\rm H}$, the equilibrium angles $\varphi_{\rm P}$ and $\varphi_{\rm F}$ of the magnetization vectors can be obtained. 

The magnetization dynamics in our system is described by the Landau-Lifschitz-Gilbert equation (LLG) adapted to our purpose:
\begin{equation}
\frac{d\mathbf{M}_{i}}{dt} = - \gamma (\mathbf{M}_{i} \times H_{i}) + \frac{\alpha_i}{M_S}(\mathbf{M}_{i} \times \frac{d\mathbf{M}_{i}}{dt})
\end{equation}
with $i$ = F, P.
One should expect that each layer follows independently this equation, hence in angular coordinates it can be expressed as:
\begin{align}
\begin{split}
\frac{d\theta_i}{dt} =& \frac{\gamma}{(1+\alpha_i^2)} 
								 (H_{\varphi_i} + \alpha_i H_{\theta_i} )  \\ 
\sin\theta_i \frac{d\varphi_i}{dt} =& \frac{\gamma}{(1+\alpha_i^2)} 
							      (\alpha_i H_{\varphi_i} - H_{\theta_i} ) 
\label{eq:LLG}
\end{split}
\end{align}
where $H_{\varphi}$ and $H_{\theta}$ are the azimuthal and polar components of the effective field,  $\alpha$ the dimensionless damping parameter and $\gamma$ is the gyromagnetic ratio, which is the same for both layers.
The effective field components can be expressed as: 
\begin{align}
\begin{split}
H_{\theta_i} =& -\frac{1}{M_S} \frac{\partial E}{\partial \theta_i} 
					     + \mathbf{h}^{\rm rf} \cdot \hat{\theta}_i  \\
H_{\varphi_i} =& -\frac{1}{M_S \sin \theta_i} \frac{\partial E}{\partial \varphi_i} 
					     + \mathbf{h}^{\rm rf} \cdot \hat{\varphi}_i
\label{eq:Heff}
\end{split}
\end{align}
where $\mathbf{h}^{\rm rf}$ is the dynamic component of the applied external field, $\hat{\theta}_i = \cos \varphi_i \cos \theta_i \hat{x} + \sin \varphi_i \cos \theta_i \hat{y} - \sin \theta_i \hat{z}$ and  $\hat{\varphi}_i = - \sin \varphi_i \hat{x} + \cos \varphi_i \hat{y}$

In our specific case $\sin\theta_{\rm P} = \sin\theta_{\rm F} = 1$, which means that we are able to rewrite Eq.\ref{eq:LLG}, for the F and P layers, as follows:
\begin{equation}
\begin{bmatrix}
\dot{\theta}_{\rm P} \\
\dot{\varphi}_{\rm P} \\
\dot{\theta}_{\rm F} \\
\dot{\varphi}_{\rm F}
\end{bmatrix}
=
-\frac{\gamma}{M_S}
\boldsymbol{[\Lambda]}
\begin{bmatrix}
\partial E / \partial \theta_{\rm P} \\
\partial E / \partial \varphi_{\rm P} \\
\partial E / \partial \theta_{\rm F} \\
\partial E / \partial \varphi_{\rm F}
\end{bmatrix}
+
\gamma
\boldsymbol{[\Lambda]}
\begin{bmatrix}
\mathbf{h}^{\rm rf} \cdot \hat{\theta}_{\rm P} \\
\mathbf{h}^{\rm rf} \cdot \hat{\varphi}_{\rm P} \\
\mathbf{h}^{\rm rf} \cdot \hat{\theta}_{\rm F} \\
\mathbf{h}^{\rm rf} \cdot \hat{\varphi}_{\rm F}
\end{bmatrix}
\label{eq:mat1}
\end{equation}
where
\begin{equation}
\boldsymbol{[\Lambda]} = 
\begin{bmatrix}
\frac{1}{1+\alpha^2_{\rm P}}
\begin{pmatrix}
\alpha_{\rm P} & 1 \\
-1 & \alpha_{\rm P} \\
\end{pmatrix}
& 0 \\
0 &
\frac{1}{1+\alpha^2_{\rm F}}
\begin{pmatrix}
\alpha_{\rm F} & 1 \\
-1 & \alpha_{\rm F} \\
\end{pmatrix}
\end{bmatrix}
\end{equation}

\subsection{Susceptibility tensor}

The differential susceptibility tensor $\boldsymbol{[\chi]} = d\mathbf{M}/d\mathbf{H}$ characterizes the dynamic magnetic response of as system to an external field. In our system, $\boldsymbol{[\chi]}$ gives us the relation between the radio frequency field $\mathbf{h}^{\rm rf}$ and the oscillating part $\dot{\mathbf{M}}$ of the magnetization vectors 
\begin{equation}
\dot{\mathbf{M}}_{\rm F} +\dot{\mathbf{M}}_{\rm P} = \boldsymbol{[\chi]} \dot{\mathbf{h}}^{\rm rf}
\end{equation}
where
$\dot{\mathbf{M}}_i =
M_S [\sin \theta_i \dot{\theta}_i \hat{\theta}_i + 
\dot{\varphi}_i \hat{\varphi}_i]$
for each layer.
In order to give an equivalent expression in function of the angular coordinates, we define $\mathbf{\Omega} =  (\theta_{\rm P}, \; \varphi_{\rm P}, \; \theta_{\rm F}, \; \varphi_{\rm F})$, with $\mathbf{\Omega} = \mathbf{\Omega}_0 + \delta \mathbf{\Omega}^{\rm rf}$, where $\delta \mathbf{\Omega}^{\rm rf} \propto e^{j \omega t}$ are the small deviations around the equilibrium positions $\mathbf{\Omega}_0$.
The magnetization deviations $\delta \mathbf{\Omega}^{\rm rf}$ are driven by the radio frequency field $\mathbf{h}^{\rm rf}$, thus they oscillate at the same frequency $\omega$.
The projections $\mathbf{h}^{\rm rf}_\mathbf{\Omega} = \mathbf{h}^{\rm rf} \cdot \mathbf{\hat{\Omega}_0}$ are related to the magnetization oscillations by a pseudo susceptibility tensor $\boldsymbol{[X]}$ defined by:
\begin{equation}
\delta \mathbf{\Omega}^{\rm rf} = \boldsymbol{[X]}  \mathbf{h}^{\rm rf}_{\mathbf{\Omega}}
\label{eq:Osc}
\end{equation}
then, if we expand the energy terms around $\mathbf{\Omega}_0$ as in $\partial E/\partial\Omega^{\rm rf}_k = \sum_l \frac{\partial^2E}{\partial\Omega_k \partial\Omega_l} \delta\Omega^{\rm rf}_l$, Eq. \ref{eq:mat1} can be expressed as:
\begin{equation}
-\frac{\gamma}{M_S} \boldsymbol{[\Lambda]}\boldsymbol{[E_{\Omega\Omega}]}\boldsymbol{[X]} \mathbf{h}^{\rm rf}_{\mathbf{\Omega}}
+ \gamma \boldsymbol{[\Lambda]} \mathbf{h}^{\rm rf}_{\mathbf{\Omega}}
= j \omega \boldsymbol{[X]} \mathbf{h}^{\rm rf}_{\mathbf{\Omega}}
\label{eq:Chi1}
\end{equation}
where the matrix $\boldsymbol{[E_{\Omega\Omega}]}$ has elements ${E_{\Omega\Omega}}_{kl} = \frac{\partial^2E}{\partial\Omega_k \partial\Omega_l}$.
For our particular system, the non-zero values of $\boldsymbol{[E_{\Omega\Omega}]}$ are:
\begin{align}
\begin{split}
E_{\theta_{\rm P}\theta_{\rm P}}   =& M_S [4 \pi M_S - {\rm H}_{\perp}  + {\rm H} \cos(\varphi_{\rm H} -\varphi_{\rm P}) \\
									& + {\rm H_k^{P}} \cos^2\varphi_{\rm P} + {\rm H_{EB}} \cos\varphi_{\rm P} + {\rm H_R} \\
									& + {\rm H_1^J}\cos(\varphi_{\rm P} - \varphi_{\rm F}) \\
									& -2{\rm H_2^J}\cos^2(\varphi_{\rm P} - \varphi_{\rm F}) ]\\
E_{\varphi_{\rm P}\varphi_{\rm P}} =& M_S [ {\rm H} \cos(\varphi_{\rm H} - \varphi_{\rm P})  \\
									& + {\rm H_k^{P}} \cos(2 \varphi_{\rm P}) + {\rm H_{EB}} \cos\varphi_{\rm P} + {\rm H_R} \\
									& + {\rm H_1^J}\cos(\varphi_{\rm P} - \varphi_{\rm F}) \\
									&  - 2{\rm H_2^J}\cos(2(\varphi_{\rm P} - \varphi_{\rm F})) ]\\ 
E_{\theta_{\rm F}\theta_{\rm F}}  = & M_S [4 \pi M_S - {\rm H}_{\perp} + {\rm H} \cos(\varphi_{\rm H}-\varphi_{\rm F}) \\
									& + {\rm H_k^{F}} \cos^2 \varphi_{\rm F} \\
									& + {\rm H_1^J}\cos(\varphi_{\rm P} - \varphi_{\rm F}) \\
									& -2{\rm H_2^J}\cos^2(\varphi_{\rm P} - \varphi_{\rm F}) ]\\
E_{\varphi_{\rm F}\varphi_{\rm F}} =& M_S [ {\rm H} \cos(\varphi_{\rm H} - \varphi_{\rm F}) + {\rm H_k^{F}} \cos(2 \varphi_{\rm F}) \\
									& + {\rm H_1^J}\cos(\varphi_{\rm P} - \varphi_{\rm F}) \\
									& - 2{\rm H_2^J}\cos(2(\varphi_{\rm P} - \varphi_{\rm F})) ]\\ 
E_{\theta_{\rm P}\theta_{\rm F}} =& \,E_{\theta_{\rm F}\theta_{\rm P}} \\
								 =& -M_S[{\rm H_1^J} + 2{\rm H_2^J}\cos(\varphi_{\rm P} - \varphi_{\rm F})]\\
E_{\varphi_{\rm P}\varphi_{\rm F}} =& \,E_{\varphi_{\rm F}\varphi_{\rm P}} \\
								   =& -M_S[{\rm H_1^J}\cos(\varphi_{\rm P} - \varphi_{\rm F}) \\
								    & + 2{\rm H_2^J}\cos(2(\varphi_{\rm P} - \varphi_{\rm F}))]
\end{split}
\end{align}
where ${\rm H_R}$ and ${\rm H}_{\perp}$ are the effective rotatable and perpendicular anisotropy fields.
For an arbitrary field $\mathbf{h}^{\rm rf}$ oscillating at a frequency $\omega$ we can obtain the pseudo susceptibility tensor using:
\begin{equation}
\boldsymbol{[X]} = \left(j \frac{\omega}{\gamma} \boldsymbol{[\Lambda]}^{-1} 
+ \boldsymbol{[E_{\Omega\Omega}]} \right)^{-1}
\end{equation}
This equation can be efficiently solved by standard numerical methods, resulting in the susceptibility tensor for each applied external field ${\rm H}$ and excitation frequency $\omega$. 

\subsection{Resonant Frequencies}

One of the important features in our systems are the resonant frequencies. These can be obtained from:
\begin{equation}
\frac{\gamma}{M_S} \boldsymbol{[\Lambda]}\boldsymbol{[E_{\Omega\Omega}]} \delta \mathbf{\Omega}
= - j \omega_r \delta \mathbf{\Omega}
\label{eg:eigen}
\end{equation}
This equation can be solved as an eigensystem using numerical methods. The eigenvalues provide us the resonant frequencies $\omega_r$ and the eigenvector values of $\delta \mathbf{\Omega}$ at that frequency.
Two positive values of $\omega_r$ are found for each external field $\rm H$.
The acquired values of $\delta \mathbf{\Omega} = (\delta \theta_{\rm P} \; \delta \varphi_{\rm P} \; \delta \theta_{\rm F} \; \delta \varphi_{\rm F})$ show that the amplitude of the out-of-plane oscillations is negligible, i. e., almost zero ($\delta \theta_{\rm P} \approx  \delta  \theta_{\rm F} \approx  0$) as expected.
The analysis of the in-plane oscillations $\delta \varphi_{\rm P}$ and $\delta \varphi_{\rm P}$ of a given eigenvector allows us to determine which is the most oscillating layer at the frequency of the corresponding eigenvalue.
When $|\delta \varphi_{\rm P}| > |\delta  \varphi_{\rm F}|$, we associate the obtained eigenvalue to the natural resonant frequency $\omega_{\rm P}$ of the pinned layer.
The opposite case ($|\delta \varphi_{\rm P}| < |\delta  \varphi_{\rm F}|$) is associated to the natural frequency $\omega_{\rm L}$ of the free layer.
One must notice that $\delta \mathbf{\Omega}$ values obtained by this method are multiplied by an unknown amplitude and phase, thereafter they are not suitable for calculating the absorbed power or to compare them at different fields ${\rm H}$. However, they provide relevant information on the relative phase and amplitude of oscillation of each layer over the dispersion relation.

\subsection{Absorbed Power}
\begin{figure}[h!]
\includegraphics{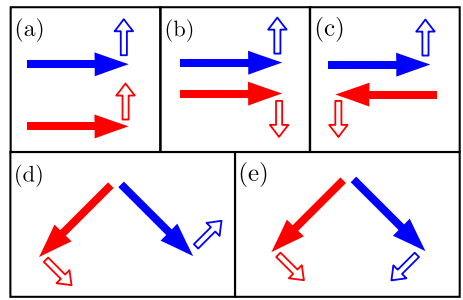}
 \caption {\label{fig:Vectors}Geometrical representation of the magnetization vectors (filled arrows) and oscillating vectors (empty arrows). Magnetizations are shown in parallel states (a and b), anti-parallel state (c) and noncollinear states (d and e). The oscillations are in phase in a, c and d or out of phase in b and e. If $\mathbf{h}^{\rm rf}$ is along the vertical direction, a and e should have larger absorbed power than b c or d
 }
\end{figure}

In order to compare directly our calculations to the experimental results, it is important to write down the average power absorbed by our system at a given field and frequency. To proceed, we start describing the instant power, per unit of volume, absorbed by our system:
\begin{equation}
P = - \mathbf{h}^{\rm rf} \cdot (\dot{\mathbf{M}}_{\rm F} + \dot{\mathbf{M}}_{\rm P})
\label{eq:PInst}
\end{equation}
It must be noticed that the amplitude of $P$ depends on three factors: (a) the relative orientation between $\mathbf{h}^{\rm rf}$ and the oscillating vectors $\dot{\mathbf{M}}_i$; (b) the temporal phase difference between $\dot{\mathbf{M}}_{\rm F}$ and $\dot{\mathbf{M}}_{\rm P}$; (c) the relative orientation of the oscillating vectors that depends on the direction of the magnetization at the equilibrium position for each layer. A graphical representation of several possible cases are presented in Fig. \ref{fig:Vectors}. 

The average absorbed power, per unit of volume, over an oscillatory cycle will also depends on the temporal phase difference between $\mathbf{h}^{\rm rf}$ and the magnetization response, which can be calculated by:
\begin{align}
\begin{split}
< P >  =& - \omega M_S \, {\rm Im} \!  \left[ \sum_{p} {{\rm h}^{\rm rf}_{\Omega}}_p \delta \Omega^{\rm rf}_q \right] \\
  =& - \omega M_S \, {\rm Im} \! \left[ \sum_{p,q} {{\rm h}^{\rm rf}_{\Omega}}_p X_{pq} {{\rm h}^{\rm rf}_{\Omega}}_q \right] 
\end{split}
\label{eq:PAve}
\end{align}

\section{Discussion}
\label{sec:Discussion}

Here we separate the discussion in samples that exhibit coupling between the ferromagnetic layers and the sample with uncoupled layers. The coupling strength were obtained by comparing the calculated and experimental data. All the simulation parameters are resumed in table \ref{Tab.I}

\begin{table}[h!]
\caption{Parameters used in simulations.}
\label{Tab.I}
\begin{ruledtabular}
\begin{tabular}{lccc}
\multicolumn{4}{c}{Common parameters to all samples} \\
\hline
$M_S$  $(\rm emu/cm^3)$ & 800 \cite{JApplPhys.105.113914} & $\rm H^{F}_K \, (Oe)$ & 5\\
$\gamma$  $(\rm MHz/Oe)$ & 17.59 \cite{JApplPhys.105.113914} & $\rm H^{P}_K \, (Oe)$ & 15 \\
$\alpha_{\rm Pinned}$ & 0.018 & $\rm H_R \, (Oe)$ & 9 \\
$\alpha_{\rm Free}$ & 0.010 & &\\
\hline
\multicolumn{4}{c}{Sample dependent parameters} \\
$t_{\mathrm{Cu}}$ & 2.5\,nm & 1.0\,nm & 0.75\,nm \\
\hline
$\rm H_{EB} \, (Oe)$        & $70$ & $83$ & $81$ \\
$\rm H^J_1 \, (Oe)$         & $0$ & $13$ & $35$ \\
$\rm H^J_2 \, (Oe)$         & $0$ & $1.0$ & $4.5$ \\
$\rm H_{\perp} \, (Oe)$  	& $0$ & $600$ & $600$ \\
$\varphi_{\rm H}$           & $4^{\circ}$ & $2^{\circ}$ & $5^{\circ}$ \\
\end{tabular}
\end{ruledtabular}
\end{table}
We choose the damping constants values in such a way that they reproduce the field widths observed at 3.7 \, GHz (see Fig. \ref{fig:AbsPeaks}). No frequency dependence of the damping parameters were considered in this work.

\subsection{No coupling}

When there is no coupling between the FM layers, our system behaves as two independent systems. The hysteresis loop can be treated as the sum of two square loops, one (centered) corresponding to F and the other one field shifted by ${\rm H_{EB}}$ corresponding to P.
The broadband response is also the addition of the individual response of each layer.
In our model, the matrix $\boldsymbol{[E_{\Omega\Omega}]}$ is then formed by two independent block of matrices along the main diagonal. Thus, an independent solution can be found for each block, corresponding to the F and P layers of our samples.
The solutions for the resonant frequencies, when the damping is neglected, are the well-known Kittel relations:
\begin{align}
\begin{split}
\omega^P_r =& \gamma \sqrt{4\pi M_S - {\rm H}_{\perp} \pm {\rm H} \pm {\rm H_{EB}} + {\rm H_k^P} + {\rm H_R}}  \\
			& \times \sqrt{\pm {\rm H} \pm {\rm H_{EB}} + {\rm H_k^P} + {\rm H_R}} \\
\omega^L_r =& \gamma \sqrt{4\pi M_S - {\rm H}_{\perp} \pm {\rm H} + {\rm H_k^F}} \sqrt{\pm {\rm H} + {\rm H_k^F}} 
\end{split}
\end{align}
 the $\pm$ sign should be chosen accordingly to the direction of the respective magnetic layer, $+$ for $\varphi=0$ and $-$ for $\varphi=\pi$ corresponding to the right or left resonant branches experimentally observed.
The resonant branches cross each other when the layers are in the anti-parallel state and where the external field is
\begin{equation}
\rm H = H_0 - \frac{1}{2} (H_{EB} + H_k^P + H_R - H_k^F)
\label{eq:Ho}
\end{equation}
At this point, the total absorbed power is the sum of the individual absorbed powers of each layer.

\subsection{Coupled FM layers}

\begin{figure}[h!]
\includegraphics{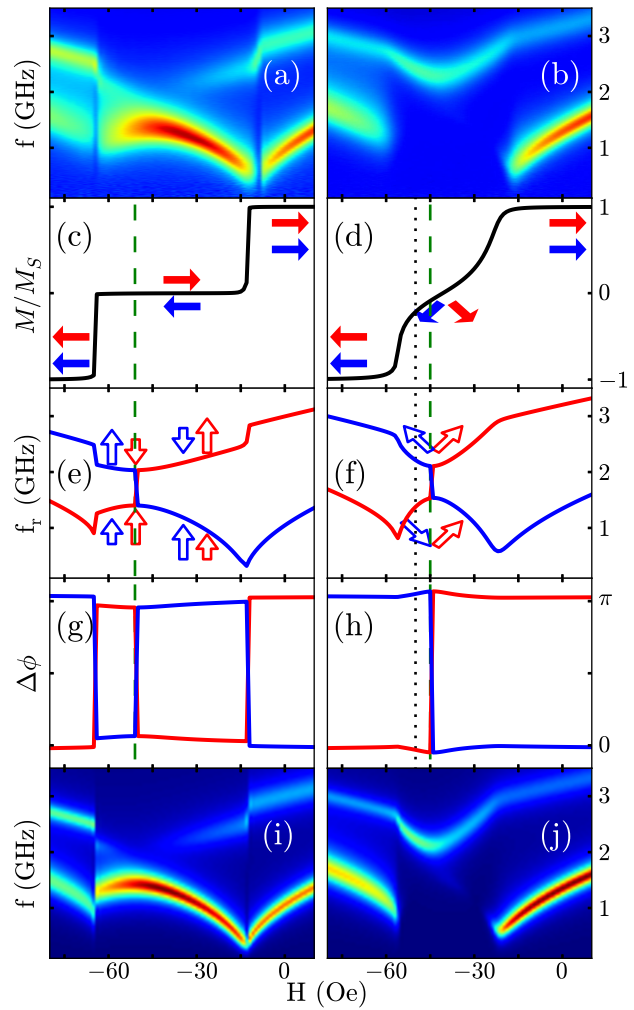}
 \caption {\label{fig:ExpSim} Experimental and simulation details in the non-saturated states for $t_{\mathrm{Cu}}$ = 1.0\,nm (left) and $t_{\mathrm{Cu}}$ = 0.75\,nm (right).
Experimental broadband spectra (a and b), simulated magnetization curves (single branch) (c and d), simulated dispersion relation (e and f), relative phase between the oscillations (g and h) and simulated broadband average absorbed power (i and j).
Filled arrows represent the magnetization vectors and empty arrows represent the oscillating vectors.}
\end{figure}

In this subsection, Fig. \ref{fig:ExpSim} resumes both the experimental and numerical results in a expanded H scale. In this figure, the colors were assigned red and blue for the pinned and free layers, respectively. Also, the filled arrows represent the magnetization vectors whilst empty arrows represent the oscillating vectors.

Having Fig. \ref{fig:ExpSim} in mind, we realize that the hysteresis response of the F layer is no longer centered. For a positive bilinear interaction (${\rm J_1} > 0$), the loops of each F layer are field shifted toward the position of the P layer loop. If the coupling is not large, as for the sample with  $t_{\mathrm{Cu}}$ = 1.0\,nm, the square shape the loops is maintained, indicating that the magnetization flips between the parallel and anti-parallel direction with respect to the external field. For the sample with $t_{\mathrm{Cu}}$ = 0.75\,nm, whose coupling intensity is larger, the hysteresis loop is no longer square, but instead acquires a rounded shape due the simultaneous rotation of the FM layers (see Fig. \ref{fig:ExpSim}(d)).

When excited by an external rf field, the magnetization of each layer does not oscillate independently. Instead, they oscillate coherently but with correlated amplitude and phase difference. In this case, we find that for a resonant mode with frequency $\omega_r$, the phase difference of the oscillations $\Delta\phi = arg[\delta \varphi_{\rm F}] - arg[\delta \varphi_{\rm P}]$ depends on the natural frequency of the companion layer: a natural frequency higher or lower than $\omega_r$ gives rise to a phase difference $\phi \approx 0{^\circ}$ or $\phi \approx 180{^\circ}$ for each case, respectively. 
This behavior has the effect of changing the the resonant peak amplitudes as seen in Fig. \ref{fig:AbsPeaks}.


Independently of the coupling intensity, there is an external field value $\rm H_0$ where both layers oscillate at the same frequency (see the dashed line in Fig. \ref{fig:ExpSim}). As long as the anti-parallel state holds, which is the case of $t_{Cu}$ = 1.0 nm, this field takes the same value as in the uncoupled case (see Eq.\ref{eq:Ho}).
In all other cases, as for $t_{Cu}$ = 0.75 nm, the magnetization angles must be taken into account for calculating $\rm H_0$, giving a complicate analytical expression. The resulting values are, however, usually close to the former cases (see the dotted line on the right panel of Fig. \ref{fig:ExpSim}).

At this field $\rm H_0$, two resonant frequencies, rather than one as in the uncoupled case, are found. The gap between these frequencies is proportional to the intensity of the given coupling. At this magnetization state, both layers oscillate with the same amplitude ($|\delta \varphi_{\rm P}| =  |\delta  \varphi_{\rm F}|$) at any given frequency. This may indicates that the dominant interaction at this state is the coupling energy. 

From Fig. \ref{fig:ExpSim}(a-b), we notice two arc-shaped (lower and upper) branches over the non saturated regime. These features can be reproduced by our model, which results in the dispersion relations and the simulated broadband average absorbed power shown in Fig. \ref{fig:ExpSim}.
Besides, our model allow us to identify that these branches are formed by both oscillating modes, corresponding to the F and P layers respectively.
The frequency gap between the branches correspond to the frequency jump at $\rm H_0$.

We calculated the oscillating vectors by taking into account both the magnetization state and the relative phase $\Delta\phi$ (shown in Fig. \ref{fig:ExpSim}(g-h)).

Interestingly, for $t_{\mathrm{Cu}}$ = 1.0\,nm, the absorption in the lower arc is favoured when the oscillating vectors are in the same direction. In this case, the oscillating vectors are also in the same direction of the rf field.
On the other hand, the sample with $t_{\mathrm{Cu}}$ = 0.75\,nm shows an appreciable absorption only in the upper inverted arc. Here the oscillating vectors (calculated at $\rm H_0$) are no longer parallel to the rf field. Instead, their vector sum is nearly parallel to the rf field for the upper branch and nearly perpendicular to the rf field for the lower arc. 

\section{Conclusions} 

In summary we reported the broadband resonance spectra in coupled and uncoupled magnetic layers in a single spin-valve configuration, namely  NiFe(20\,nm)/Cu($t_{\mathrm{Cu}}$)/NiFe(20\,nm)/IrMn(15\,nm) where $t_{\mathrm{Cu}}$ = 0.75\,nm, 1.0\,nm and 2.5\,nm controls the coupling intensity.
For coupled cases, we observed that, at low field, the experimental broadband spectra is complex, while at high field the spectra show the typical behavior of coupled saturated samples. The coupling between the ferromagnetic layers was observed to modify the relative amplitudes of the absorption peaks.

We were able to reproduce remarkably well the broadband experimental results, both in saturated and non-saturated states, by using our numerical method based on the macro spin approximation, obtaining the dispersion relations from Eq.\ref{eg:eigen} and the broadband average absorbed power from Eq.\ref{eq:PAve}.
The method provide further insights on the magnetization dynamics in coupled systems, predicting frequency gaps and complex dispersion relations in non-saturated magnetic states. Such states, besides of their importance in applications, are usually neglected by the traditional descriptions of both regular FMR and broadband FMR experiments.

As final comments we would like to point that our matrix mathematical approach allows any one to easily describe magnetic systems with an arbitrary number of interacting macro spins. It also allows us an easy and fast software implementation of the method by using well established numerical subroutines \cite{scipy}.

\begin{acknowledgments}
The authors thank to Dr. Marcio Assolin Corr\^{e}a for the fruitful discussion and revision.
This work has been supported by the Brazilian agencies CNPq, FINEP, FAPERJ and CAPES.
\end{acknowledgments}

\bibliography{Refs2}

\end{document}